\documentclass{iaus}
\usepackage{graphicx}

\title[IAUS254.~~The Galactic Disk-Halo Transition] 
{The Galactic Disk-Halo Transition \\ -- Evidence from Stellar Abundances
\thanks{Based on observations made with the ESO Telescopes at the Paranal
observatory under programs 65.L-0507, 67.D-0439, 68.D-0094, 68.B-0475,
69.D-0679, 70.D-0474, and 76.B-0133.}}

\author[Poul E. Nissen \& William J. Schuster]   
{Poul Erik Nissen$^1$
 \and William J. Schuster$^2$}

\affiliation{$^1$Department of Physics and Astronomy, University of Aarhus, DK-8000
Aarhus C, Denmark \\ email: {\tt pen@phys.au.dk} \\[\affilskip]
$^2$Observatorio Astron\'{o}mico Nacional, Universidad Nacional Aut\'{o}noma
de M\'{e}xico, Apartado Postal 877, 
Ensenada, BC, 22800 M\'{e}xico, \\email: {\tt schuster@astrosen.unam.mx}}

\pubyear{2008}
\volume{254}  
\pagerange{16}
\jname{The Galaxy Disk in Cosmological Context}
\editors{J. Andersen, J. Bland-Hawthorn \& B. Nordstr\"{o}m, eds.}
\begin{document}

\maketitle

\begin{abstract}
New information on the relations between the Galactic disks, the halo,
and satellite galaxies is being obtained from elemental abundances of
stars having metallicities in the range $-1.5 <$ [Fe/H] $< -0.5$.
The first results for a sample of 26 halo stars and 13 thick-disk stars
observed with the ESO VLT/UVES spectrograph are presented.
The halo stars fall in two distinct groups:
one group (9 stars) has [$\alpha$/Fe]= $0.30 \pm 0.03$ like the thick-disk stars.
The other group (17 stars) shows a clearly deviating trend 
ranging from [$\alpha$/Fe]= 0.20 at [Fe/H]= $-1.3$ to [$\alpha$/Fe]= 0.08 at
[Fe/H]= $-0.8$. The kinematics of the stars are discussed and the abundance
ratios Na/Fe, Ni/Fe, Cu/Fe and Ba/Y are applied to see if the
``low-alpha'' stars are connected to the thin disk
or to Milky Way satellite galaxies. Furthermore, we compare our data 
with simulations of chemical abundance
distributions in hierarchically formed stellar halos in a $\Lambda$CDM
Universe.
\end{abstract}

\firstsection 
\section{Introduction}

As discussed by Venn et al. (2004), the chemical signatures of stars in Milky Way
dwarf spheroidal (dSph) galaxies are different from the majority of Galactic thin-disk,
thick-disk and halo stars. The dSph stars have lower values of $\alpha$/Fe, where
$\alpha$ refers to the abundance of typical alpha-capture elements like Mg, Si, Ca and Ti,
than the Galactic stars. Furthermore, the Ba/Y ratio is significantly higher in the dSph
stars. Thus, Venn et al. conclude that the main components of our Galaxy have not
been formed through the merging of dwarf galaxies similar to present-day dSphs. 
This could be seen as an argument against hierarchical structure formation as
predicted in Cold Dark Matter cosmologies. Font et al. (2006) have, however, made
$\Lambda$CDM simulations of the chemical abundance distributions 
in the the Galactic halo, which
show that early accreted dwarf galaxies have chemical properties different from those
of present-day satellite galaxies. For the metal-poor part of the Galactic halo, they
predict enhanced values of [$\alpha$/Fe] corresponding to the ratio calculated 
for Type II supernovae, but for the metal-rich end of the halo a
decline of [$\alpha$/Fe] is predicted due to an increasing contribution of iron
from Type Ia SNe.

In order to test these recent simulations, we have started a survey of abundance ratios
of about 100 halo stars having $-1.5 <$ [Fe/H] $< -0.5$.
This metallicity range
also contains thin-disk and thick-disk stars, and hence a comparison of abundances 
may give new information on the relations between the various Galactic populations.

The survey is a continuation of a more limited study by Nissen \& Schuster (1997), 
who determined abundances of 13 halo stars and 16 thick disk stars in the
metallicity range $-1.3 <$ [Fe/H] $< -0.4$. Interestingly,
8 of the halo stars turned out to have significantly lower values of
[$\alpha$/Fe]\footnote{Defined as 
[$\alpha$/Fe] = $\frac{1}{4}$\,\,([Mg/Fe]\,+\,[Si/Fe]\,+\,[Ca/Fe]\,+\,[Ti/Fe])}
than the other halo stars. Considering the 
small number of halo stars studied it is, however, unclear from the work of
Nissen \& Schuster if the distribution
of [$\alpha$/Fe] in the halo is continuous or bimodal. One of the aims of the 
present study is to answer this question by observing a larger sample of stars.

\section{Selection of stars and observations}
The stars have been selected from Schuster et al. (2006), which contains
$uvby$-$\beta$ photometry and complete kinematic data for 1533 high-velocity and metal-poor
stars. In order to ensure that a selected star belongs to the halo population, we
required that its total space velocity with respect to the Local Standard of Rest (LSR),
$V_{\rm total}$, should be larger than 180\,km\,s$^{-1}$. 
Furthermore, we used the Str\"{o}mgren indices, $b-y$,
$m_{\rm 1}$ and $c_{\rm 1}$, to select dwarfs and subgiants with temperatures 
$5100 < T_{\rm eff} < 6200$\,K and metallicities $-1.5 <$ [Fe/H] $< -0.5$.
When selecting the stars from the bright end of the Schuster et al. catalogue, these
criteria result in a limiting magnitude of $V = 11.1$ for stars that could be 
reached with the Nordic Optical Telescope (NOT) on La Palma. 

The first observations
with NOT and its FIbre fed Echelle Spectrograph (FIES) were carried out
in May 2008 and resulted in high resolution ($R \sim 45\,000$) spectra with 
$S/N \sim 150$ for 30 halo stars. The abundance analysis of these data is, however,
not yet finished. Instead, we are presenting here results for 26 halo stars
that fulfil our selection criteria, and for which spectra obtained with
the VLT/UVES spectrograph are available in the ESO science archive.
These spectra have $R \sim 60\,000$ and very high signal-to-noise ratios, $S/N > 300$,
in the 4800 - 6500\,\AA\ region that we are using. 
In addition, 13 thick-disk stars with UVES spectra are included. Their atmospheric
parameters fall in the same ranges as the halo stars, but their total space velocities
relative to the LSR are typical for thick-disk stars 
($50 < V_{\rm total} < 150$\,km\,s$^{-1}$).

\section{Abundance analysis}
Abundances of the elements Mg, Si, Na, Ca, Ti, Cr, Ni, Cu, Y and Ba are derived from
equivalent widths of weak to medium-strong lines ($5 < EW < 80$\,m\AA ). A
grid of 1D MARCS model-atmospheres is used to derive the abundances, and
the effect on the electron pressure from variations of [$\alpha$/Fe] is taken
into account. Like 
Nissen \& Schuster (1997), the analysis is performed in a differential way with
respect to two bright thick-disk stars, HD\,22879 and HD\,76932, that are
known to have [$\alpha$/Fe] close to +0.3\,dex according to several studies (e.g. Reddy et al.
2006). For nearby stars that appear to be unreddened, as judged from the absence of interstellar
NaD lines, $T_{\rm eff}$ is derived from the $b-y$ and $V-K$ color indices, and
the gravity parameter, log$g$, is estimated via the Hipparcos parallax. For the more
distant and reddened stars, $T_{\rm eff}$ is determined from the excitation balance of Fe\,{\sc i}
lines, and log$g$ is derived from the requirement that the difference in Fe abundances
derived from Fe\,{\sc ii} and Fe\,{\sc i} lines should  be the same as in HD\,22879 and
HD\,76932. The adopted metallicity, [Fe/H], is that derived from Fe\,{\sc ii}
lines.

\begin{figure}[ht]
\begin{center}
 \includegraphics[width=13.0cm]{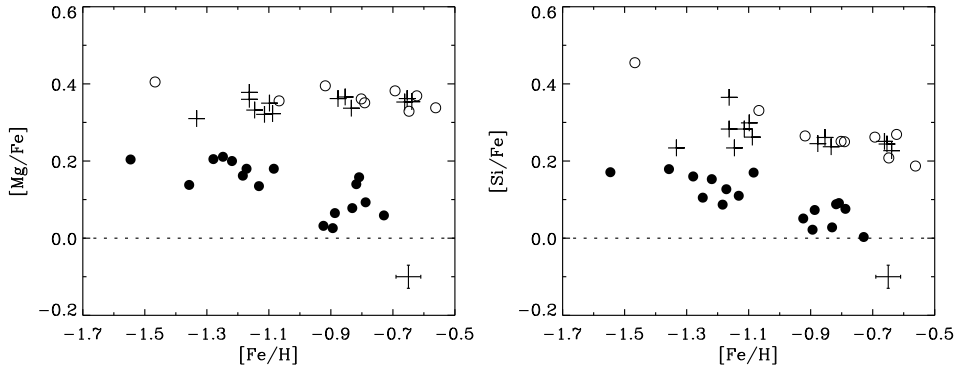}
 \caption{[Mg/Fe] and [Si/Fe] vs. [Fe/H] for the sample of stars with VLT/UVES spectra.
Crosses: Thick-disk stars; Open circles:  ``High-alpha'' halo stars;
Filled circles:  ``Low-alpha'' halo stars. Typical error bars for the data are shown in the
lower right corners of the figures.}
   \label{fig1}
\end{center}
\end{figure}

\begin{figure}[ht]
\begin{center}
 \includegraphics[width=13.0cm]{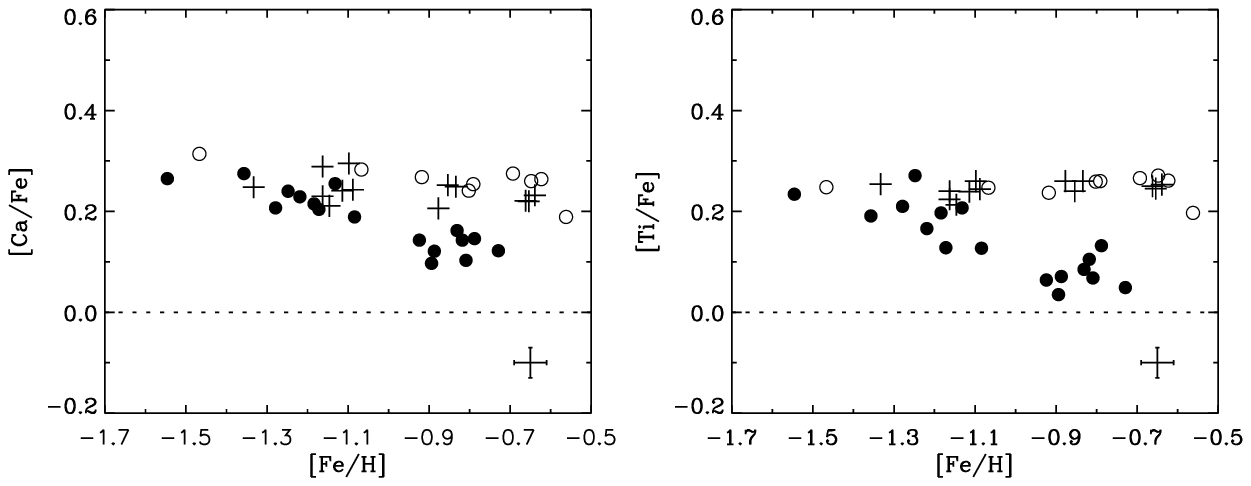}
 \caption{[Ca/Fe] and [Ti/Fe] vs. [Fe/H]. Same symbols as in Fig. 1.}
   \label{fig2}
\end{center}
\end{figure}

In order minimize the dependence of abundance ratios on possible errors in the atmospheric
parameters, the abundance ratio of two elements is derived from lines belonging
to the same ionization stage,
e.g. [Mg/Fe] from Mg\,{\sc i} and Fe\,{\sc i} lines or
[Ba/Y] from Ba\,{\sc ii} and Y\,{\sc ii} lines. Local thermodynamic equilibrium
(LTE) is assumed, but due to the limited
range of the parameters of our stars, non-LTE effects on the
derived {\em differential} abundances are expected to be small.

\section{Results and discussion}
Figures  1 and 2 show the derived abundances of the four alpha-capture
elements, Mg, Si, Ca and Ti, relative to Fe. As seen, the halo stars fall
in two distinct groups: $i)$ the ``high-alpha'' stars that have very near
the same [$\alpha$/Fe] as the thick-disk stars and a remarkably small
scatter, $\pm 0.03$\,dex, around the ``plateau'' value of [$\alpha$/Fe]
and $ii)$ the ``low-alpha'' stars that show a
decreasing trend of [$\alpha$/Fe] as a function of increasing [Fe/H].
In the cases of Mg and Si the separation of the two groups of
halo stars is seen for the whole range $-1.5 <$ [Fe/H] $< -0.7$ with a 
maximum separation of $\sim 0.25$\,dex in [Mg/Fe] and $\sim 0.20$\,dex
in [Si/Fe]. For Ca and Ti the two groups tend to merge at [Fe/H] $\simeq -1.2$,
and the maximum separation is only about 0.12\,dex for Ca and about 0.17\,dex
for Ti. 

\begin{figure}[ht]
\begin{center}
 \includegraphics[width=13.0cm]{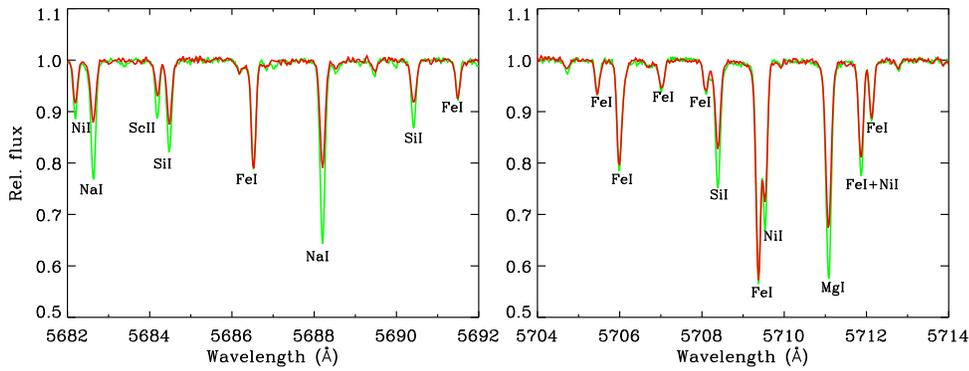}
 \caption{Spectra of two stars with nearly the same atmospheric parameters
$T_{\rm eff}$, log$g$ and [Fe/H].  The spectrum of the ``low-alpha" star CD\,$-45\,3283$
($T_{\rm eff}$\,=\,5603\,K, log$g$\,=\,4.57, [Fe/H]\,=\,$-0.89$, [$\alpha$/Fe]\,=\,0.08)
is shown with a thick (red) line, and that of the ``high-alpha" star G\,159-50
($T_{\rm eff}$\,=\,5648\,K, log$g$\,=\,4.39, [Fe/H]\,=\,$-0.92$, [$\alpha$/Fe]\,=\,0.29)
with a lighter (green) line.}
   \label{fig3}
\end{center}
\end{figure}

\begin{figure}[ht]
\begin{center}
 \includegraphics[width=13.0cm]{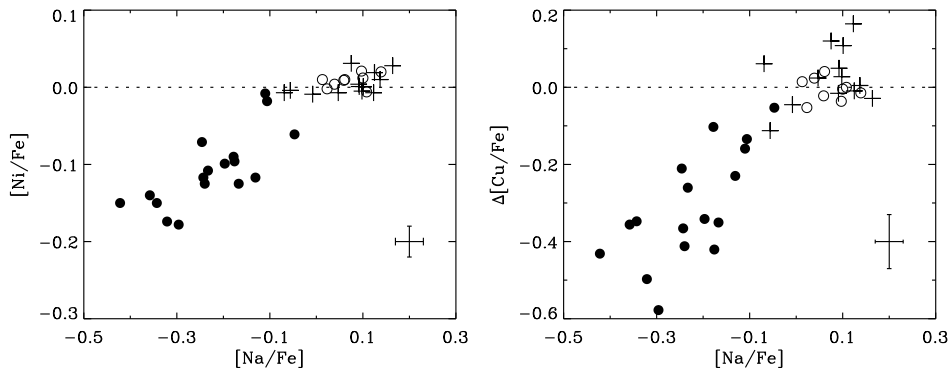}
 \caption{[Ni/Fe] and [Cu/Fe] vs. [Na/Fe] with the same symbols as in Fig. 1.
In the case of Cu the deviation of [Cu/Fe] from the relation defined by the
thick-disk stars is plotted.}
   \label{fig4}
\end{center}
\end{figure}

A possible explanation of these trends is that the ``high-alpha'' and 
the thick-disk stars have been formed in regions with a relative high
star-formation rate such that only Type II SNe have contributed to
the chemical evolution up to [Fe/H] $\sim -0.5$. The ``low-alpha''
stars, on the other hand, come from regions with a slower star-formation
rate, where Type Ia SNe have started to contribute with iron at a
metallicity [Fe/H] $\simeq -1.5$. The reason for the smaller separation
in [Ca/Fe] compared to [Mg/Fe], may be that Mg is almost
exclusively produced in Type II SNe, whereas about 25\% of Ca
originates in Type Ia SNe according to the chemical evolution models
of Tsujimoto et al. (1995).

The abundance differences between the two halo groups can be seen
directly from the observed spectra. Fig. 3 shows two spectral regions for a
``high-alpha'' and a ``low-alpha'' star with similar $T_{\rm eff}$, log$g$
and [Fe/H] values. As seen, the Fe\,{\sc i} lines of the two stars have
nearly the same strength, whereas the Mg and Si lines are
weaker in the ``low-alpha'' star. The same is the case for Na, Ni and
Cu lines and, as shown in Fig. 4, [Ni/Fe] and [Cu/Fe] are well correlated 
with [Na/Fe]. The reason for these correlations may be that the yields of
Na, Ni and Cu depend on the neutron excess in supernovae and that this excess
is affected by the $\alpha$/Fe ratio.

\begin{figure}[ht]
\begin{center}
 \includegraphics[width=10.0cm]{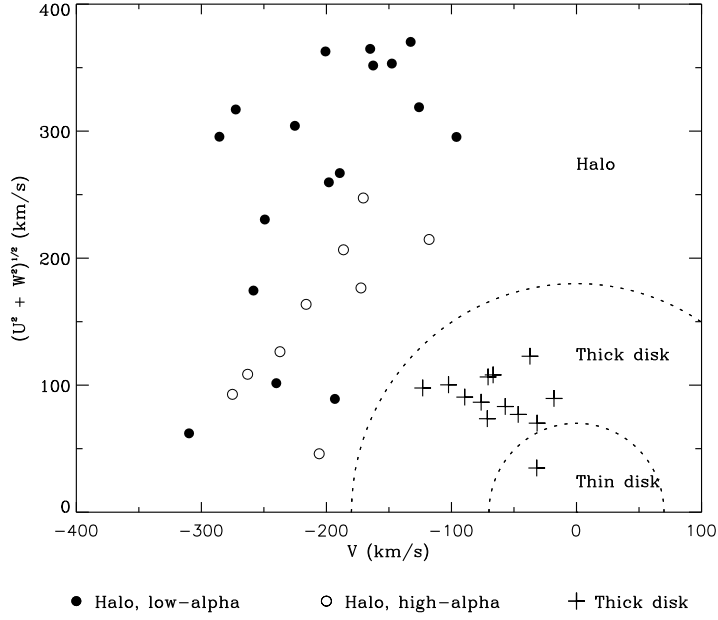}
 \caption{The ``Toomre'' diagram for the three groups of stars discussed
in the present paper. Regions, where stars are most likely to belong
to the halo, the thick disk and the thin disk, respectively
(see Venn et al. 2004, Fig. 1), are separated
by dashed circles corresponding to $V_{\rm total}$\,=\,180 and 70\,km\,s$^{-1}$.}
   \label{fig5}
\end{center}
\end{figure}

\begin{figure}[ht]
\begin{center}
 \includegraphics[width=7.0cm]{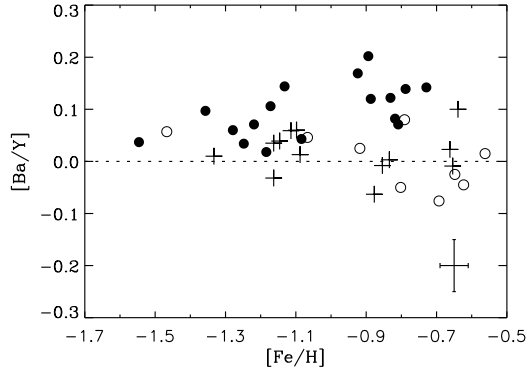}
 \caption{[Ba/Y] vs. [Fe/H] with the same symbols as in Fig. 1}
   \label{fig6}
\end{center}
\end{figure}

The kinematics of the stars are shown in Fig. 5. As seen from this ``Toomre''
diagram, both groups of halo stars have an average Galactic rotation velocity close
to zero in contrast to the thick-disk stars. Furthermore, the velocity
dispersion for the ``low-alpha'' group is higher than that of the 
``high-alpha'' group. This larger dispersion is mainly caused by larger $U$
velocities, which means that the ``low-alpha'' stars tend to move on higher energy
radial orbits. A larger sample of halo stars is, however, needed
before any definitive conclusions about the kinematics of the two groups
can be made.

The distribution of the ``low-alpha'' stars in Figs. 1 and 2 shows a 
smooth continuation of the $\alpha$/Fe trend for thin-disk stars,
which according to several studies (Reddy et al. 2003,
Bensby et al. 2005) stops with [$\alpha$/Fe] $\simeq 0.1$ at [Fe/H] $\simeq -0.7$.
The thin-disk stars do not share, however, the low [Na/Fe], [Ni/Fe] and [Cu/Fe] values
that are being found for the ``low-alpha'' group. In addition, the ``low-alpha''
stars have enhanced Ba/Y ratios as shown in Fig. 6.
In all these abundance deviations, the ``low-alpha'' stars resemble 
stars in dSph galaxies (Venn et al. 2004, Sbordone et al. 2007, Koch et al. 2008)
and in the LMC (Pomp\'{e}ia et al. 2007),
although these satellite galaxies tend to have larger abundance offsets from disk stars
than the ``low-alpha'' stars.

The trend of [Mg/Fe] for the ``low-alpha'' halo stars in Fig. 1 agrees remarkably
well with the trend predicted by Font et al. (2006, Fig. 9) from simulations of 
abundance distributions for a hierarchically formed stellar halo in a 
$\Lambda$CDM Universe. The fact that the ``low-alpha'' stars do not have quite
as low [$\alpha$/Fe] values as present-day satellite galaxies also agrees with
their predictions. On the other hand, the ``high-alpha'' halo stars are
not predicted from the simulations. The existence of this group suggests
that the formation of the Galactic halo is more complicated than predicted
from the $\Lambda$CDM simulations. It remains to be seen if this group of ``high-alpha'' halo
stars can be explained as due to the merger of an exceptionally large satellite or
if it is a dissipative component of the Galaxy as suggested by Gratton et al. (2003).
In the latter case one would expect the group of ``high-alpha'' stars to have some
net rotation. Another problem is that the ``low-alpha'' halo stars tend
to move on high-energy radial orbits that plunge into the Galactic central
regions from the outer regions of the halo. As discussed by Gilmore \& Wyse (1998),
this requires possible parent satellite galaxies to have a very high mean density
to provide ``low-alpha'' stars on orbits with such small perigalactic distances.

In order to obtain more insight into these problems, the elemental
abundances of a larger sample of metal-rich halo stars should be studied so
that better knowledge of the kinematics and 
the relative frequency of halo stars belonging to the ``high-alpha'' and
``low-alpha'' groups can be obtained.
As mentioned earlier, such data is being obtained with the Nordic 
Optical Telescope.

\medskip

{\em Acknowledgement}. This work has been financially supported
from CONACyT project 49434-F.

\end{document}